\documentclass[trackchanges,twocolumn]{aastex62}
\usepackage{amsmath,amstext}
\usepackage{units}
\usepackage{url}
\usepackage{xspace}

\newcommand{\E}[1]{\times10^{#1}}
\newcommand{\msol}{ \, M_\sun}

\newcommand{\bi}{\begin{itemize}}
\newcommand{\ei}{\end{itemize}}
\newcommand{\commentOut}[1]{}
\newcommand{\sedona}{\texttt{Sedona}\xspace}
\newcommand{\cmfgen}{\texttt{CMFGEN}\xspace}	

\shortauthors{Shen et al.}

\begin{document}

\title{\bf \Large{Non-Local Thermodynamic Equilibrium Radiative Transfer Simulations of Sub-Chandrasekhar-Mass White Dwarf Detonations}}

\author[0000-0002-9632-6106]{Ken J.\ Shen}
\affiliation{Department of Astronomy and Theoretical Astrophysics Center, University of California, Berkeley, CA 94720, USA}

\author[0000-0002-9388-2932]{St\'{e}phane Blondin}
\affiliation{Unidad Mixta Internacional Franco-Chilena de Astronom\'{i}a, CNRS/INSU UMI 3386 and Instituto de Astrof\'isica, Pontificia Universidad Cat\'olica de Chile, Santiago, Chile}
\affiliation{Aix Marseille Univ, CNRS, CNES, LAM, Marseille, France}

\author{Daniel Kasen}
\affiliation{Department of Astronomy and Theoretical Astrophysics Center, University of California, Berkeley, CA 94720, USA}
\affiliation{Department of Physics, University of California, Berkeley, CA 94720, USA}
\affiliation{Lawrence Berkeley National Laboratory, Berkeley, CA, USA}

\author[0000-0003-0599-8407]{Luc Dessart}
\affiliation{Institut d'Astrophysique de Paris, CNRS-Sorbonne Universit\'{e}, 98 bis boulevard Arago, F-75014 Paris, France}

\author[0000-0002-9538-5948]{Dean M.\ Townsley}
\affiliation{Department of Physics \& Astronomy, University of Alabama, Tuscaloosa, AL, USA}

\author{Samuel Boos}
\affiliation{Department of Physics \& Astronomy, University of Alabama, Tuscaloosa, AL, USA}

\author[0000-0001-5094-8017]{D.\ John Hillier}
\affiliation{Department of Physics and Astronomy \& Pittsburgh Particle Physics, Astrophysics, and Cosmology Center (PITT PACC), University of Pittsburgh, Pittsburgh, PA 15260, USA}

\correspondingauthor{Ken J. Shen}
\email{kenshen@astro.berkeley.edu}

\begin{abstract}

Type Ia supernovae (SNe~Ia) span a range of luminosities and timescales, from rapidly evolving subluminous to slowly evolving overluminous subtypes.  Previous theoretical work has, for the most part, been unable to match the entire breadth of observed SNe~Ia with one progenitor scenario.  Here, for the first time, we apply non-local thermodynamic equilibrium radiative transfer calculations to a range of accurate explosion models of sub-Chandrasekhar-mass white dwarf detonations.  The resulting photometry and spectra are in excellent agreement with the range of observed non-peculiar SNe~Ia through $\unit[15]{d}$ after the time of $B$-band maximum, yielding one of the first examples of a quantitative match to the entire \cite{phil93a} relation.  The intermediate-mass element velocities inferred from theoretical spectra at maximum light for the more massive white dwarf explosions are higher than those of bright observed SNe~Ia, but these and other discrepancies likely stem from the one-dimensional nature of our explosion models and will be improved upon by future non-local thermodynamic equilibrium radiation transport calculations of multi-dimensional sub-Chandrasekhar-mass white dwarf detonations.

\end{abstract}


\section{Introduction}

Type Ia supernovae (SNe~Ia) are the explosions of C/O white dwarfs (WDs) in interacting stellar systems (see \citealt{maoz14a} and \citealt{jha19a} for recent reviews).  Their standardizable light curves, powered by the radioactive decay of $^{56}$Ni \citep{pank62a,cm69}, can be observed from nearly halfway across the Universe, enabling their use as cosmological distance indicators \citep{ries98,perl99}, and their thermonuclear ashes contribute significantly to the metal content of galaxies \citep{tww95}. However, despite scrutiny dating back to at least 185 AD and a sharp rise in understanding over the past few decades, fundamental uncertainties remain as to the identity of the companion star(s) and the means by which the companion ignites the exploding WD.

One such explosion mechanism, the double detonation of a sub-Chandrasekhar-mass WD, in which a helium shell detonation triggers a carbon core detonation, has been explored as a potential SN~Ia explosion model beginning four decades ago \citep[e.g.,][]{nomo82b,wtw86}.  However, in this early work, the helium shells were relatively massive ($\sim 0.1 \msol$), leading to large amounts of thermonuclear ash on the outside of the ejecta, which yielded light curves and spectra that were inconsistent with observations \citep[e.g.,][]{nuge97}.

More recently, it was realized that both stable and unstable mass transfer in double WD systems can lead to double detonations with far smaller helium shells that yield much better agreement with observations of SNe~Ia \citep{bild07,fhr07,guil10,pakm13a}.  Using a more realistic nuclear reaction network, which includes additional reaction pathways that are neglected in standard networks with a limited number of isotopes, \cite{shen14b} found that helium shell ashes for these lower-mass envelopes are dominated by Si and Ca and would thus produce observables that better agree with observations (perhaps even explaining the high-velocity features seen in most SNe~Ia).  This result has been confirmed by recent multi-dimensional simulations \citep{town19a,gron20a}.

These studies, combined with significant issues with other SN~Ia scenarios (see \citealt{maoz14a} and \citealt{jha19a} and references therein), have made it increasingly likely that sub-Chandrasekhar-mass double detonations in double WD systems give rise to a fraction, if not the majority, of non-peculiar SNe~Ia, ranging from subluminous to overluminous subtypes.  In fact, the strongest evidence to date for a successful SN~Ia scenario came with \emph{Gaia}'s second data release, in the form of hypervelocity stars that could only have been realistically produced by the disruption of double WD binaries \citep{shen18b}.  This is a natural outcome of double detonations in double degenerate systems undergoing dynamically driven mass transfer (the D$^6$ scenario) for which the explosion of the sub-Chandrasekhar-mass primary WD may occur before the complete tidal disruption of the secondary WD.  While only three such stars were found, they may represent the brightest examples of a much larger underlying population of D$^6$ survivors in the Solar neighborhood.

These successes have spurred an increasing interest in generating detailed observables of sub-Chandrasekhar-mass WD detonations in order to compare to observed SNe~Ia \citep{sim10,blon17a,blon18a,shen18a,gold18a,poli19a,town19a,gron20a,kush20a}.  These studies simulate the exploding WD in a range of detail, from one-dimensional models of bare C/O WDs to multi-dimensional calculations with realistic helium shells.  However, all but \cite{blon17a,blon18a} have been performed assuming that energy level populations of ions and, in most cases, ionization state populations are given by local thermodynamic equilibrium (LTE), which becomes an increasingly inaccurate approximation as SNe evolve.  In this work, we present non-LTE radiative transfer calculations of a suite of one-dimensional bare C/O WD explosions with accurate nucleosynthesis.  We find that radiation transport calculations using a combination of non-LTE and more realistic explosion models yield excellent photometric and spectroscopic agreement with observations, further supporting the idea that sub-Chandrasekhar-mass WD detonations can reproduce the full range of observed SN Ia properties.  We do find some discrepancies with observations, but these will likely be overshadowed by the changes to predicted observables when calculations with realistic helium shells are performed in multiple dimensions.


\section{Hydrodynamical explosion models}

The starting models for our radiative transfer calculations are one-dimensional solar metallicity WD explosions from \cite{shen18a}.  In that study, artificially broadened detonations were ignited at the center of the WDs and followed with the reactive hydrodynamic code, \texttt{FLASH} \citep{fryx00}, augmented with a 41-isotope nuclear network using the \texttt{MESA} \citep{paxt11} nuclear reaction module with rates from JINA's \texttt{REACLIB} \citep{cybu10a}.  Tracer particles were then post-processed with a 205-isotope nuclear network, again using \texttt{REACLIB} and \texttt{MESA}'s infrastructure.  No mixing prescriptions are applied.  In this work, we use the $0.85$, $0.90$, $1.00$, and $1.10 \msol$ simulations from \cite{shen18a} with uniform C/O mass ratios of $50/50$ and $30/70$ for a total of eight models.  Their explosion properties are listed in Table \ref{tab:prop} in the Appendix.  We note that C/O mass ratios in WDs are likely lower than 50/50 (\citealt{giam18a} use asteroseismology to infer a C/O ratio as low as 20/80 for one WD), but we include the 50/50 models for comparison to previous literature.

The nucleosynthetic yields from these models are in good agreement with recent simulations of bare C/O WD detonations that take care to treat the unresolved detonation realistically such as \cite{mile19a} and \cite{kush20a}.  However, they disagree significantly with past studies  including \cite{sim10} and \cite{blon17a}, due primarily to the older studies' less accurate implementation of detonation physics and/or oversimplification of the nuclear network.  The discrepancies are particularly large for the lower mass explosions, for which the previous $^{56}$Ni masses are as much as a factor of $\sim 3$ lower than those for the models we use.  The agreement is better at higher masses $\gtrsim 1.0 \msol$ because more mass is processed through nuclear statistical equilibrium, which is relatively insensitive to the choice of nuclear network and the treatment of the detonation.  The nucleosynthetic differences are particularly important given the tantalizing correspondence to the observed \cite{phil93a} relation that \cite{blon17a} found for masses $ \gtrsim 1.0 \msol$ using the non-LTE radiative transfer code \cmfgen.  Since these high-mass models are in agreement with more accurate explosion calculations, there is hope that non-LTE radiative transfer of a wide range of accurate models will yield a better match to the entirety of SNe~Ia, from subluminous to overluminous subtypes.


\section{Light curves}

To this end, we process our eight explosion models with the radiative transfer code \sedona and a subset of these models with \cmfgen.  We focus on non-LTE calculations in this Letter, but in the Appendix, we detail results using various LTE prescriptions in \sedona to inform past and future LTE calculations.


\subsection{\sedona snapshots and the impact of non-LTE}
\label{sec:nlte}

\sedona \citep{ktn06} is a Monte Carlo radiative transfer code that has been used in many studies of astronomical transient phenomena.  Most of these studies have assumed that energy level and ion state populations possess their LTE values, but these can also be calculated in non-LTE in \sedona.  Due to computational constraints, in this Letter, we apply \sedona's non-LTE capabilities to ``snapshot'' calculations, in which photons are propagated through ejecta whose velocity, density, and composition profiles are fixed at a particular time.  This is done iteratively until the output luminosity matches that of a time-dependent LTE calculation and radiative equilibrium in each zone is established.  Further details are outlined in Section \ref{sec:sednlte}.

\begin{figure}
  \centering
  \includegraphics[width=\columnwidth]{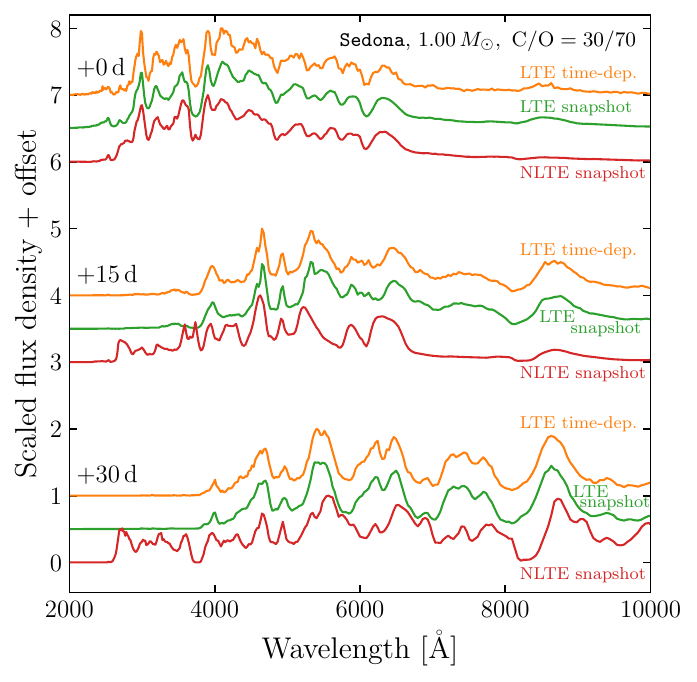}
  \caption{Scaled and offset \sedona spectra of a $1.0 \msol$, ${\rm C/O}=30/70$ WD detonation $+0$, $+15$, and $\unit[+30]{d}$ from $B$-band maximum.  Time-dependent and snapshot LTE spectra using the resolved line opacity formalism described in Sec.\ \ref{sec:bb} are shown in orange and green, respectively, and non-LTE snapshot spectra are shown in red.}
  \label{fig:spec_100_3070_sedona}
\end{figure}

To confirm that this snapshot procedure has the potential to match cost-prohibitive time-dependent non-LTE calculations, we compare spectra of time-dependent and snapshot LTE  simulations in Figure \ref{fig:spec_100_3070_sedona} for the $1.0 \msol$ 30/70 C/O model; models with different masses and C/O mass ratios behave similarly.  It is clear that the LTE snapshot spectra are nearly identical to the time-dependent LTE spectra at the three sampled times.  We thus proceed with the non-LTE snapshots under the assumption that these would also match time-dependent non-LTE calculations.

As can be seen in Figure \ref{fig:spec_100_3070_sedona}, the maximum-light LTE and non-LTE spectra are very similar in the optical, confirming previous work by \cite{baro96a}.  However, differences between our LTE and non-LTE calculations are apparent by day $+15$, predominantly due to a hotter line-forming region in non-LTE, a higher ratio of Fe \textsc{iii} to Fe \textsc{ii}, less blanketing by Fe \textsc{ii} lines at bluer wavelengths, and a commensurate decrease in the redistribution of flux at redder wavelengths.  This has important implications for post-maximum photometry as well as the \cite{phil93a} relation, which we discuss in Section \ref{sec:phil}.


\subsection{Time-dependent non-LTE with \cmfgen and comparison to observations}
\label{sec:cmfgen}

\begin{figure*}
  \centering
  \includegraphics[width=\textwidth]{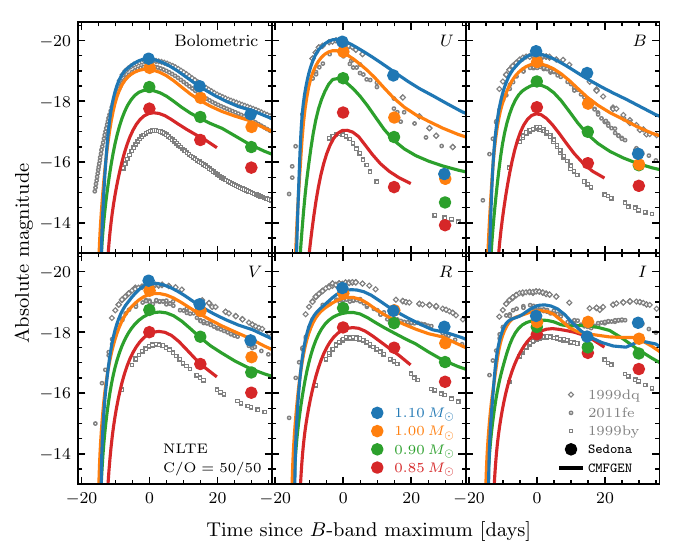}
  \caption{Multi-band non-LTE light curves of 50/50 C/O mass ratio WD detonations.  Masses are as labeled, with brighter peak $B$-band magnitudes as the mass increases.  \cmfgen results are represented by solid lines; \sedona snapshots are demarcated with circles.  Open diamonds, circles, and squares show observations of SNe~1999dq, 2011fe, and 1999by, respectively.  Note that the observed ``bolometric'' light curves only consider UVOIR flux and are thus up to $\sim 20\%$ lower than the true bolometric luminosities.}
  \label{fig:lcs_5050}
\end{figure*}

We also perform time-dependent radiative transfer calculations with the non-LTE code \cmfgen \citep{hill12a,dess14b} for a subset of the models: all four of the 50/50 C/O models and the $1.0 \msol$ 30/70 C/O model.  The Appendix contains a more complete description of \cmfgen, but we note here the significant fact that the atomic data input for the two codes is not identical.  In particular, the \cmfgen runs consider more levels for Fe \textsc{ii}, Co \textsc{ii}, and Co \textsc{iii} than are included in the \sedona calculations.  Thus, comparisons of results from the two codes in this study should be regarded as qualitative.

Figure \ref{fig:lcs_5050} shows light curve comparisons of the \sedona non-LTE snapshot and the \cmfgen non-LTE time-dependent calculations for our models with ${\rm C/O}=50/50$; our ${\rm C/O}=30/70$ models are displayed in the Appendix in Figure \ref{fig:lcs_3070}.  As with the \sedona LTE calculations, there is good agreement at the time of $B$-band maximum in all bands (e.g., the two codes yield maximum $B$-band magnitudes that differ by $\unit[0.2]{mag}$ at most; see Tab.\ \ref{tab:prop} in the Appendix for more details), and the bolometric, $V$-band, and $R$-band light curves continue to match through day $+30$.  Moreover, the agreement for most bands and models persists through day $+15$.  Given the different methods of solving the radiative transfer problem in \sedona and \cmfgen, this agreement gives confidence that the multi-band photometry is being accurately calculated in non-LTE through day $+15$.

As shown in Figures  \ref{fig:lcs_5050} and \ref{fig:lcs_3070},  the agreement through day $+15$ also extends to the observed photometry of the overluminous SN~1999dq \citep{stri05a,jha06b,gane10a}\footnote{Most of the observational data used in this work was obtained through \texttt{https://sne.space} \citep{guil17a}.}, which is well-matched in most bands by the $1.1 \msol$ explosions; the normal SN~2011fe \citep{muna13b,pere13a,tsve13a}, which corresponds to the $1.0 \msol$ detonations; and the subluminous SN~1999by \citep{garn04,stri05a,gane10a}, which agrees with the $0.85 \msol$ explosions.  There are some discrepancies: for example, the $R$- and $I$-band light curves for the $1.1 \msol$ explosion are too dim compared to SN~1999dq, the rise times of the more massive models are too short compared to observations, and the colors at very early times for all the models are too red.  But overall, the agreement through day $+15$ for this wide range of SNe~Ia is striking.

The \cmfgen and \sedona light curves do diverge from each other and from observations by day $+30$ in $U$-, $B$-, and $I$-bands.  The disagreement in bluer bands is especially large for the higher masses.  Perhaps coincidentally, the photometry from the two codes brackets the best-matching observed $U$-band and $B$-band light curves at day $+30$, with \cmfgen being too bright and \sedona being too faint.

The root cause of the disagreement between the two codes is unclear.  As previously discussed, the input atomic data is not identical for the two codes, rendering quantitative comparisons difficult.  Another possibility is \sedona's assumption of radiative equilibrium (i.e., that heating is equal to cooling in each zone), which is not the case for \cmfgen.  \cmfgen also includes non-thermal processes, which are neglected in the \sedona runs.  These only contributed tenths of a magnitude of difference at day $+30$ for the Chandrasekehar-mass WD models in \cite{dess14b}, but their effects may be stronger for sub-Chandrasekhar-mass WD explosions.  As it is beyond the current scope, we leave a thorough examination of the causes of differences between the two codes to a future study.


\subsection{The Phillips relation}
\label{sec:phil}

\begin{figure}
  \centering
  \includegraphics[width=\columnwidth]{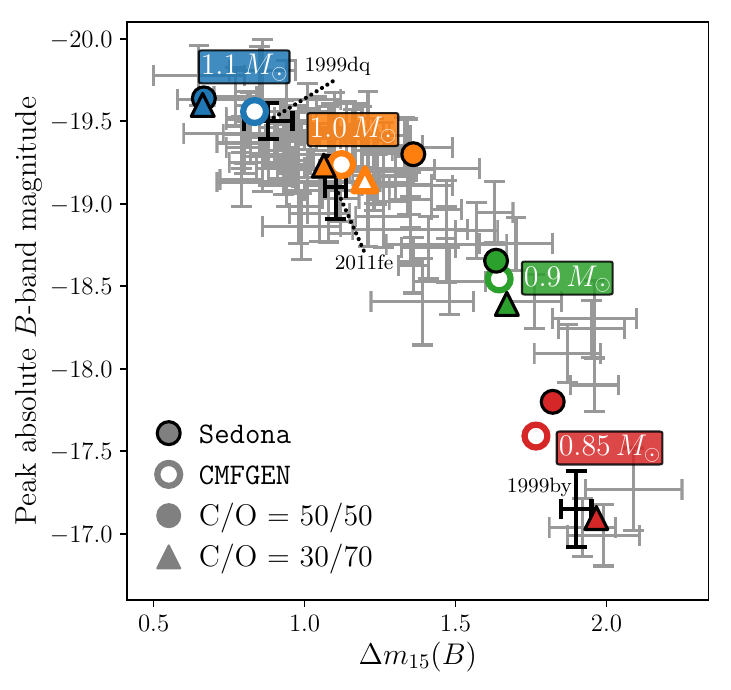}
  \caption{Peak $B$-band magnitude vs.\ decrease in $B$-band magnitude between peak and $\unit[15]{d}$ after peak: the \cite{phil93a} relation.  Solid symbols show the results from the \sedona non-LTE snapshots, and open symbols show the \cmfgen results.  Circles (triangles) represent initial C/O fractions of 50/50 (30/70).  Masses are as labeled.  Gray error bars are values from a sample of SNe~Ia, and black error bars are values for SN 1999by, SN 2011fe, and SN 1999dq.}
  \label{fig:phillips}
\end{figure}

Figure \ref{fig:phillips} shows the peak $B$-band magnitude vs.\ $\Delta m_{15}(B)$, the decline in $B$-band magnitude $\unit[15]{d}$ after peak, for our \sedona and \cmfgen simulations.  Gray error bars are observed values taken from the CfA light curve data set \citep{hick09a}, and black error bars represent SN 1999by \citep{garn04}, SN 2011fe \citep{pere13a}, and SN 1999dq \citep{gane10a}.

It has been well-established for decades that SNe~Ia that reach brighter peak $B$-band magnitudes evolve more slowly than dimmer SNe~Ia; this is the \cite{phil93a} relation.  Most studies of Chandrasekhar-mass explosions (but c.f.\ \citealt{hoef17a}) have been unable to reproduce the entirety of the relation due to the constant ejecta mass in the explosions, which is the primary variable controlling the rate of the light curve evolution.  These studies have found that, while the more luminous end of the relation can be achieved by Chandrasekhar-mass explosions, they cannot reproduce observations with $\Delta m_{15}(B) \gtrsim 1.5$ \citep[e.g.,][]{kase09b,sim13a,blon17a}.

Previous work on sub-Chandrasekhar-mass WD explosions \citep[e.g.,][]{sim10,blon17a,shen18a} failed to match observations simultaneously at both the bright and the dim end of the relation.  The present work  is the first study that has combined more accurate detonation modeling with non-LTE radiative transfer, and as a result, our radiative transfer simulations reproduce the entirety of the Phillips relation, from subluminous to overluminous SNe~Ia.  We note that this agreement would not be as convincing if the \cite{phil93a} relation incorporated data more than $\unit[15]{d}$ after $B$-band maximum: our \cmfgen and \sedona calculations become discrepant soon after day $+15$, with neither code adequately matching the light curves of the entire range of observed SNe~Ia.  However, it is reassuring that, when the results of the two codes do match each other, they also match observations.

Peak magnitudes and values of $\Delta m_{15}(B)$ have also been found to correlate with host galaxy properties.  \cite{shen17c} combined a binary population synthesis calculation with an assumed relationship between exploding WD masses and the resulting values of $\Delta m_{15}(B)$ to derive a quantitative explanation for this observed correlation of SN~Ia properties with host galaxy characteristics (in particular, stellar age).  The relationship we find in the present work between WD mass and $\Delta m_{15}(B)$ is consistent with that assumed in \cite{shen17c}, and thus their conclusions remain valid: namely,  sub-Chandrasekhar-mass WD detonations are a plausible mechanism to explain the correlation of  SN~Ia properties with stellar age due to the evolution of the primary WD mass in merging double WD binaries.

We also note that the relationship between peak magnitude and the rate of decline for the bolometric light curve is much weaker.  The bolometric decline rate parameter $\Delta m_{15}({\rm bolometric})$ is indeed larger for the $0.85 \msol$ models than the $1.1 \msol$ models, but only by $\unit[0.2]{mag}$.  This confirms earlier work by \cite{kase07a} that the primary driver of the $B$-band width-luminosity relation is not the photon diffusion time but instead the faster color evolution of the dimmer models due to the earlier onset of Fe \textsc{iii} to Fe \textsc{ii} recombination.


\section{Spectra}
\label{sec:spec}

\begin{figure*}
  \centering
  \includegraphics[width=\textwidth]{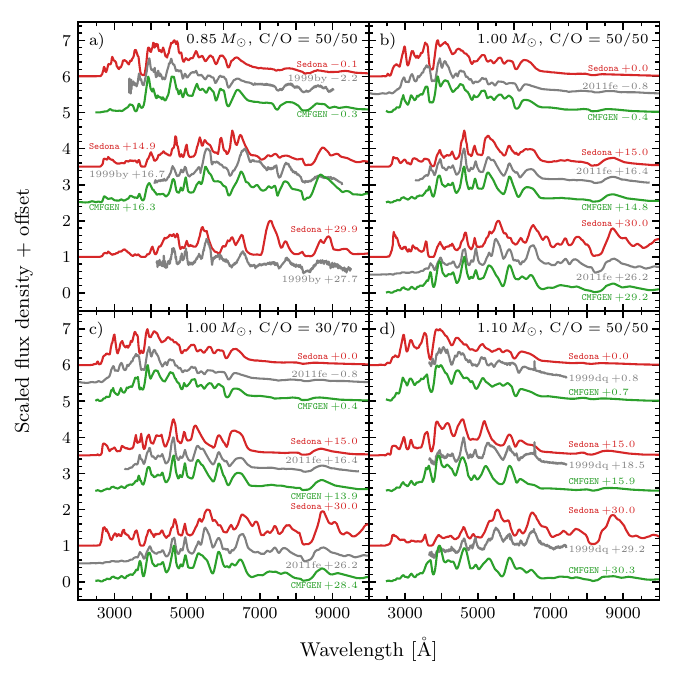}
  \caption{Scaled and offset \cmfgen and \sedona non-LTE spectra with comparisons to spectra of observed SNe with similar light curves.  Days from the time of $B$-band maximum are as labeled.}
  \label{fig:spec}
\end{figure*}

Scaled and offset \sedona and \cmfgen spectra of a subset of the models near $+0$, $+15$, and $\unit[+30]{d}$ from $B$-band maximum are shown in Figure \ref{fig:spec}, along with comparisons to observed spectra.  (The $0.9 \msol$ models are omitted for simplicity, and models without \cmfgen spectra are displayed in Fig.\ \ref{fig:spec_alt} in the Appendix.)  The near-maximum spectra are all in excellent agreement with each other, except for one significant flaw: as with previously published LTE spectra \citep[e.g.,][]{sim10,shen18a,poli19a} and those in this work, the intermediate-mass element (IME) absorption features for the more massive WD models are blueshifted by several thousand ${\rm km \, s^{-1}}$ compared to observations; see Table \ref{tab:prop} in the Appendix for the inferred velocities at maximum light.

However, the multi-dimensional study performed by \cite{town19a} yielded maximum-light spectral IME velocities for a $1.0 \msol$ explosion that are much closer to that seen in SN~2011fe  along most lines of sight.  \cite{gron20a} found a similar result for their $1.05 \msol$ explosion, albeit in an angle-averaged sense.  It is thus likely that the IME velocity discrepancies are a product of the one-dimensional nature of our models, given the multi-dimensional LTE results and the similarity between our LTE and non-LTE spectra at maximum light.

At later times, the agreement among \cmfgen, \sedona, and the observations is less definitive but still persists.  E.g., the $0.85 \msol$ \sedona day $+15$ and $+30$ models do not appear to be close fits to SN~1999by, but the 50/50 \cmfgen model at day $+14$ does adequately reproduce SN~1999by's day $+17$ spectrum.  Meanwhile, the $1.0 \msol$ day $+15$ \sedona and \cmfgen spectra for both C/O ratios provide good matches to SN~2011fe's spectrum, and at day $+30$, \sedona's spectra continue to yield a good fit, while the \cmfgen spectra appear suitable.  A similar pattern arises for the $1.1 \msol$ spectra and SN~1999dq.  We also note that the maximum-light IME velocity discrepancies are not a problem at these later times.

In summary, it appears that non-LTE radiative transfer of one-dimensional sub-Chandrasekhar-mass WD detonations is able to reproduce observed spectra of the entire range of SNe~Ia, from the subluminous SN~1999by to the overluminous SN~1999dq, up to $\unit[30]{d}$ from the time of $B$-band maximum.  The most notable discrepancy is in the IME velocities at peak light for the brighter SNe~Ia, but at later times, this difference disappears.


\section{Conclusions}

Here, we summarize the main results of this work:
\bi
\item{The predicted photometry from non-LTE radiative transfer simulations of sub-Chandrasekhar-mass WD detonations using two different codes (\sedona and \cmfgen) matches through day $+15$  after $B$-band maximum for all models and in almost all bands.  Moreover, this multi-band photometry agrees relatively well with observations of a wide range of SNe~Ia through $\unit[+15]{d}$.  With minor exceptions, the theoretical bolometric, $V$-band, and $R$-band light curves match each other and observations through at least $\unit[+30]{d}$.  Some discrepancies do exist, but they will likely be overshadowed by the changes that future multi-dimensional calculations will engender.  Clumping, which we do not consider in this work, will also quantitatively change the results presented here \citep{wilk20a}.}

\item{Non-LTE radiative transfer of sub-Chandrasekhar-mass WD detonations is able to reproduce the entirety of the \cite{phil93a} relation, from subluminous to overluminous SNe~Ia.  Since the photometry generated with both \cmfgen and \sedona agrees through day $+15$, this appears to be a robust finding.}

\item{The simulated peak light spectra mostly agree with observations for the whole range of SNe~Ia, but the minima of the IME absorption features for the higher mass explosions are faster than those for the brighter SNe, whose luminosity they match.  Day $+15$ and $+30$ spectra are also reproduced (with no velocity shifts).}

\item{The predicted observables from our non-LTE simulations agree with those from our LTE calculations at the time of maximum $B$-band light.  That is to say, non-LTE effects have little visible impact at this time.  After this time, non-LTE calculations yield hotter, more ionized, and less Fe \textsc{ii}-line-blanketed atmospheres as compared to LTE simulations, leading to changes in flux in the ultraviolet and near-infrared.  Non-LTE and LTE bolometric and $V$-band light curves (and to a lesser extent, $R$-band light curves) continue to match each other until at least $\unit[+30]{d}$ from maximum.}
\ei

Our non-LTE radiative transfer calculations show that sub-Chandrasekhar-mass WD detonations match observations of the entire range of SNe~Ia, from subluminous to overluminous, perhaps as well as could be expected from one-dimensional simulations.  Radiative transfer studies of multi-dimensional double detonation simulations with low-mass helium shells are in their infancy and have only been performed in LTE \citep{town19a,gron20a}, but multi-dimensional effects have already proven important for reducing the discrepancy in IME velocities at maximum light.  However, given the work presented here, these multi-dimensional LTE light curves and spectra are likely only accurate until maximum light.  After this time, the bolometric and $V$-band light curves will probably agree with future, more physical non-LTE results, but multi-dimensional non-LTE simulations will be crucial for accurate predictions in other bands and for spectra at later times.


\acknowledgments

We thank the anonymous referee for their comments and acknowledge helpful discussions with Alison Miller.  We are indebted to the other developers of \sedona, with extra thanks to David Khatami, Hannah Klion, and Nathan Roth.  K.J.S., D.M.T., and S.B. received support for this work from NASA through the Astrophysics Theory Program (NNX17AG28G).  D.K.\ is supported in part by the U.S.\ Department of Energy, Office of Science, Office of Nuclear Physics, under contract number DE-AC02-05CH11231 and DE-SC0004658, and by a SciDAC award DE-SC0018297.  D.J.H.\ acknowledges support for the development of \cmfgen from STScI theory grant HST-AR-12640.001-A and NASA theory grant NNX14AB41G.  This research was supported in part by the Gordon and Betty Moore Foundation through grant GBMF5076, by a grant from the Simons Foundation (622817DK), and by the Exascale Computing Project (17-SC-20-SC), a collaborative effort of the U.S. Department of Energy Office of Science and the National Nuclear Security Administration.    This research used the Savio computational cluster resource provided by the Berkeley Research Computing program at the University of California, Berkeley (supported by the UC Berkeley Chancellor, Vice Chancellor for Research, and Chief Information Officer).  This research also used resources of the National Energy Research Scientific Computing Center (NERSC), a U.S.\ Department of Energy Office of Science User Facility located at Lawrence Berkeley National Laboratory, operated under Contract No.\ DE-AC02-05CH11231.


\software{\cmfgen \citep{hill12a}, \texttt{FLASH} \citep{fryx00}, \texttt{matplotlib} \citep{hunt07a}, \texttt{MESA} \citep{paxt11}, \sedona \citep{ktn06}}


\appendix


\section{Radiation transport with \sedona}

\subsection{Expansion opacity formalism}
\label{sec:eps}

For the majority of previous studies using \sedona, local thermodynamic equilibrium (LTE) is assumed to hold for the matter, so that ionization fractions and level populations are given by Saha-Boltzmann distributions.  Furthermore, most of these studies have treated line opacities in the ``expansion opacity'' approximation \citep{karp77a}, in which the contributions from bound-bound transitions are grouped together into discretized frequency bins.

In the \sedona implementation of expansion opacity, the source function for bound-bound transitions can be expressed in terms of the mean intensity integrated over the line profile, $\bar{J}_\nu$, and the Planck function, $B_\nu(T)$, as 
\begin{align}
	S_\nu = (1 - \epsilon) \bar{J}_\nu + \epsilon B_\nu(T) ,
\end{align}
where $\epsilon$ is the absorption probability, and the gas temperature is $T$.  In principle, $\epsilon$ is different for every transition of each element, but it is typically a single user-specified constant in \sedona; we adopt this simplification here, but c.f. \cite{gold18a} where it is set separately for low-mass and high-mass elements.

\begin{figure}
  \centering
  \includegraphics[width=\columnwidth]{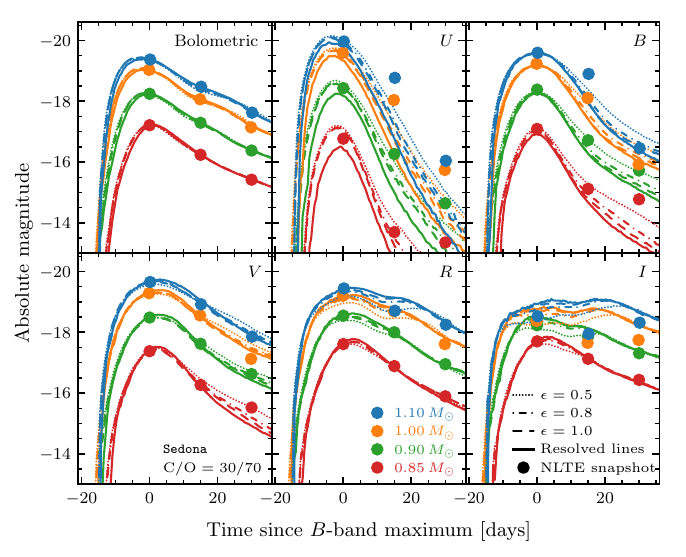}
  \caption{Multi-band \sedona light curves of ${\rm C/O}=30/70$ WD detonations with various treatments of bound-bound opacity.  Masses are as labeled, with brighter peak $B$-band magnitudes as the mass increases.  Dotted, dashed-dotted, and dashed lines show results using the LTE expansion opacity formalism (discussed in Sec.\ \ref{sec:eps}) with $\epsilon=0.5$, $0.8$, and $1.0$, respectively.  Solid lines represent calculations using resolved lines in LTE (Sec.\ \ref{sec:bb}), and non-LTE snapshots (Sec.\ \ref{sec:sednlte}) are demarcated with circles.}
  \label{fig:lcs_3070_sedona}
\end{figure}

In our previous study \citep{shen18a}, $\epsilon$ was set equal to 1.  Figure \ref{fig:lcs_3070_sedona} shows the effects of varying $\epsilon$ on LTE light curves of our sub-Chandrasekhar-mass WD detonations.  Dotted, dashed-dotted, and dashed lines represent simulations with $\epsilon=0.5$, $0.8$, and $1.0$, respectively.  (Solid lines and circles represent other treatments of bound-bound transitions and will be discussed in Secs.\ \ref{sec:bb} and \ref{sec:sednlte}.)  Bolometric light curves are largely insensitive to the choice of $\epsilon$ for the duration of the simulations, and the same is true for the light curves of each individual band except $I$-band through day $+15$ from $B$-band maximum.  After day $+15$, decreasing $\epsilon$ results in bluer light curves, with a color difference that increases with time.  The differences are relatively minimal in the $V$- and $R$-bands, but they are significant for the other bands, reaching $\unit[0.8]{mag}$ in $B$-band for the $0.85 \msol$ model at $\unit[+30]{d}$.


\subsection{Resolved line opacity formalism}
\label{sec:bb}

Line opacities can also be implemented in \sedona in a less approximate way by assuming the absorption and emission profiles are specified functions that are then directly mapped to an opacity grid.  For simplicity, we assume the two profiles are equivalent to each other.  In LTE, this implies the source function is the Planck distribution, so the effective absorption probability $\epsilon=1$.

In this work, we choose the line profile to be a Voigt profile.  If the only line broadening is thermal, resolving the lines would require an extremely finely spaced frequency grid.  We thus specify an artificial broadening velocity, set in this work to be $\unit[100]{km \, s^{-1}}$.\footnote{We verify that our results are converged with a broadening velocity as high as $\unit[200]{km \, s^{-1}}$, but we use a lower velocity of $\unit[100]{km \, s^{-1}}$ for added certainty.}  We use a logarithmic frequency spacing of $d\nu/\nu = 3\E{-5}$, yielding $\sim 10$ frequency bins per line.

The results for our ${\rm C/O}=30/70$ models using this resolved line opacity formalism assuming LTE ionization state and level populations are shown in Figure \ref{fig:lcs_3070_sedona} as solid lines.  The bolometric light curves are consistent with the expansion opacity simulations described in Section \ref{sec:eps}, and the multi-band light curves are mostly consistent with each other until maximum light, after which the resolved line opacity simulations remain somewhat close to the  expansion opacity simulations with $\epsilon = 1$.  However, in $U$-band, the resolved line opacity simulations are consistently dimmer than all of the  expansion opacity simulations.


\subsection{Non-LTE snapshots}
\label{sec:sednlte}

\sedona also allows for the inclusion of non-LTE effects in the resolved line opacity formalism by relaxing the LTE assumption that ionization state and level populations are set by Saha-Boltzmann equilibrium.  Instead, we implement non-LTE by assuming that statistical equilibrium holds, so that the time derivative of every level population is zero:
\begin{align}
	\frac{dn_{i,j,k}}{dt} &= n_{i,j+1,0}R_{c \rightarrow k} + \sum_{k'} n_{i,j,k'} R_{k' \rightarrow k} - n_{i,j,k} R_{k \rightarrow c} - n_{i,j,k} \sum_{k'} R_{k \rightarrow k'} \nonumber \\
	&= 0 ,
\end{align}
where the population of level $k$ of species $i$ in ionization state $j$ is $n_{i,j,k}$, $R_{k \rightarrow k'}$ is the transition rate from level $k$ to $k'$, $R_{c \rightarrow k}$ is the rate of transitions from the continuum ground state of ionization state $j+1$, and $R_{k \rightarrow c}$ is the transition rate from level $k$ to the continuum ground state of ionization state $j+1$.  In \sedona, this constraint is cast as a matrix inversion problem ($d \vec{n}_i/dt={\bf M} \vec{n}_i=0$) for each element that is treated in non-LTE.  In this work, we calculate the primary spectral-feature-forming elements Si, S, Ca, Fe, Co, and Ni in non-LTE and assume the rest of the elements are in LTE to reduce the computational load.

While the capability to perform time-dependent non-LTE calculations exists in \sedona, such simulations currently take too much computational time to be feasible for production runs.  Efforts are underway to accelerate these calculations, but for this study, we instead perform non-LTE snapshot calculations, in which a prescribed luminosity is propagated through the composition and density structure of a model at a given time to generate a spectrum, similar to snapshot procedures in other work (e.g., \citealt{nuge97,kerz14a}).  We note that the times of the non-LTE results are given with respect to the time of $B$-band maximum as calculated using LTE.  In principle, this time may be different for a time-dependent non-LTE calculation; however, as we describe below, predicted observables from LTE and non-LTE calculations are quite similar through the time of peak light, so this is likely a negligible difference.

To ensure that such snapshot calculations can adequately reproduce time-dependent results, we first perform snapshot calculations in LTE using the resolved line opacity formalism and compare them to the spectra produced by the time-dependent resolved line opacity LTE calculations described in Section \ref{sec:bb}; we show the results of this comparison in Figure \ref{fig:spec_100_3070_sedona}.  We use the composition and density structures at $+0$, $+15$, and $\unit[+30]{d}$ from the $B$-band maxima for all eight of our models and allow the luminosity from the radioactive decay at those times to propagate outwards.  However, the outgoing luminosity in the time-dependent calculation is not equal to the instantaneous radioactive decay power because of the finite diffusion speed of the photons as well as energy lost to expansion.  Thus, we also supplement the radioactive luminosity with additional luminosity emanating from the center of the ejecta, iterating until the total outgoing luminosity matches the value from the time-dependent calculation.  The converged value of the necessary additional luminosity is relatively insensitive to the opacity structure of the ejecta, so we only need to perform this series of iterations once for each model epoch.

The initial temperature structure from the time-dependent calculation is used to calculate opacities, and photon packets are propagated outwards until all have left the grid.  The temperature structure is then recalculated in order to yield radiative equilibrium (i.e., heating equal to cooling) in all zones.  We perform this iteration procedure 10 times for our LTE snapshots.  While 10 iterations is sufficient for convergence for all of the LTE and roughly half of the non-LTE snapshots, some of the non-LTE snapshots do not reach convergence after 10 iterations.  For these models, we continue the iteration procedure until none of their broad band magnitudes changes by more than $\unit[0.1]{mag}$ over 5 iterations.  We note that, even though the magnitudes and temperature structures do converge in all cases, there is no guarantee that they converge to the ``correct,'' physical solutions.  Due to computational restrictions, it is not feasible to redo the snapshot iteration procedure from a variety of initial conditions, which would help to ascertain how unique the converged solutions are.  A more robust methodology is to perform time-dependent non-LTE calculations, which yield much better initial conditions at each time step.  We show the results of such time-dependent simulations in Section \ref{sec:cmfgen} using \cmfgen, but at present, such calculations are too costly with \sedona.

The resulting multi-band photometry at day $+0$, $+15$, and $+30$ from $B$-band maxima for our ${\rm C/O}=30/70$ \sedona non-LTE snapshots is shown as circles in Figure \ref{fig:lcs_3070_sedona}.  Lines show the results of time-dependent LTE calculations using a variety of implementations, as described in Sections \ref{sec:eps} and \ref{sec:bb}.  By construction, the bolometric light curves are well-matched to the time-dependent resolved line opacity LTE calculations at each epoch, and, as expected from the spectral comparison, the photometry at peak light in every band is also very similar.  However, the reduction in Fe~\textsc{ii} line blanketing and commensurate reduction in Ca~\textsc{ii} emission after peak when implementing non-LTE leads to discrepancies compared to the resolved line opacity LTE calculations \citep{kase06}, with the largest differences in the $U$- $B$-, and $I$-bands: $\unit[15]{d}$ after peak, the non-LTE results are as much as $2 \,$mag brighter (dimmer) in $U$-band ($I$-band) than the resolved line opacity LTE results.  Meanwhile, the $V$-band magnitudes are relatively unchanged between LTE and non-LTE because the iron line flux redistribution mostly occurs outside of the $V$-band, leaving it largely untouched.

It is clear from this comparison that LTE radiative transfer calculations of sub-Chandrasekhar-mass SNe~Ia in most bands and under a variety of approximations are reliable until peak but not afterwards.  After $B$-band maximum, bolometric and, to a lesser extent, $V$-band light curves are relatively immune from the effects of non-LTE, but all other bands are subject to corrections as large as $\unit[2]{mag}$ just $\unit[15]{d}$ from peak.


\section{Description of \cmfgen and comparison of input data with \sedona}

\cmfgen is a non-LTE radiative transfer code that solves the time-dependent radiative transfer equations and the time-dependent kinetic equation for a homologous flow, including treatments of non-local energy deposition and non-thermal processes. No expansion-opacity formalism is used as all bound-bound transitions are explicitly included in the transfer equation. To reduce the number of bound-bound levels, super-levels are utilized. In this approach, which can be easily modified, levels in the same ion with similar properties are assumed to depart from LTE by the same factor.

We use the same setup as in \cite{blon17a,blon18a}, except for two notable differences: (a) we use a larger model atom for Fe \textsc{ii}. We consider the first 2698 full levels,\footnote{Compiled from \url{http://kurucz.harvard.edu}.} which are grouped into 228 super levels for the purpose of solving the time-dependent rate equations, whereas our previous setup consisted of 827 full levels grouped into 275 super levels; (b) the radioactive decay energy deposition function is determined assuming a purely absorptive grey opacity with $\kappa_\gamma = 0.06 \, Y_e \, {\rm cm^2 \, g^{-1}}$, as opposed to computed via a Monte Carlo code. 

The linelist used in the \sedona calculations is a superset of the linelist used in \cmfgen, with three exceptions: the \sedona linelist only includes the first 1000 levels of Fe \textsc{ii}, Co \textsc{ii}, and Co \textsc{iii}, whereas \cmfgen considers the first 2698, 2747, and 3917 levels, respectively.  This is the primary reason that comparisons of results from the two codes in this study cannot be regarded as quantitative; a proper comparison, which will be performed in the future, requires the input data to be as identical as possible, especially for these important ions.


\section{Supplemental figures and data}

\begin{figure}
  \centering
  \includegraphics[width=\textwidth]{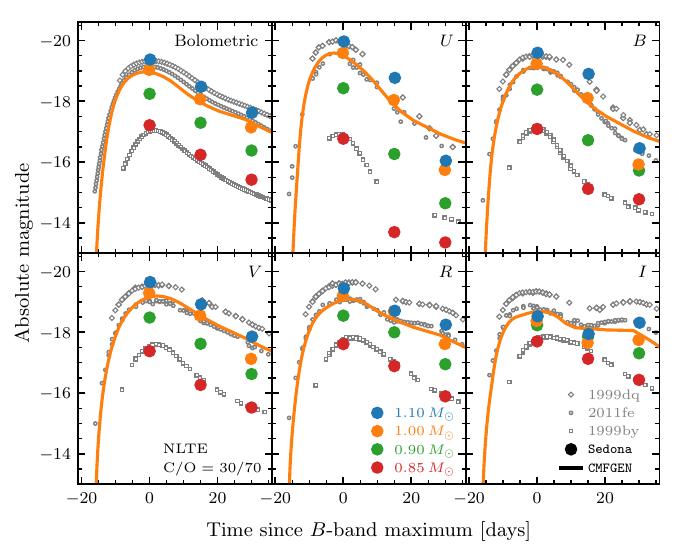}
  \caption{Same as Fig.\ \ref{fig:lcs_5050}, but for C/O mass ratios of 30/70.}
  \label{fig:lcs_3070}
\end{figure}

\begin{figure*}
  \centering
  \includegraphics[width=\textwidth]{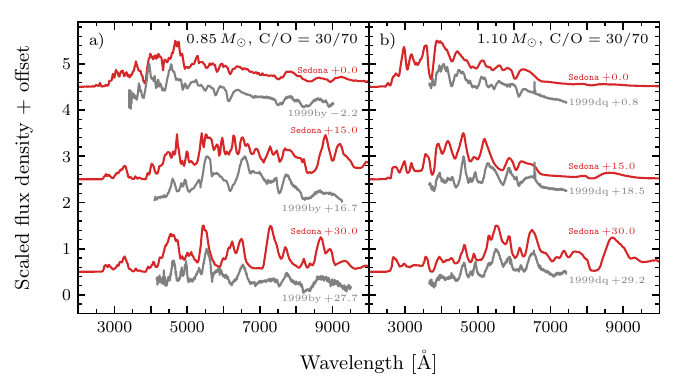}
  \caption{Same as Fig. \ref{fig:spec}, but for the two models without \cmfgen spectra.}
  \label{fig:spec_alt}
\end{figure*}

\begin{deluxetable}{c|cccc|cccc}
\tablecolumns{9}
\tablecaption{Explosion and non-LTE radiative transfer properties of the models used in this work\label{tab:prop}}
\tablehead{
\colhead{} & \multicolumn{4}{c}{${\rm C/O} = 50/50$} & \multicolumn{4}{c}{${\rm C/O}=30/70$}}
\startdata
WD mass [$M_\odot$] & 0.85  & 0.90  & 1.00 & 1.10 & 0.85  & 0.90  & 1.00 & 1.10 \\
$^{56}$Ni mass [$M_\odot$] & 0.13  & 0.28  & 0.56 & 0.80 & 0.08 & 0.23 & 0.52& 0.78 \\
Kinetic energy [$10^{51} \,$erg] & 0.87  & 1.00  & 1.22 & 1.38 & 0.65 & 0.82 & 1.08 & 1.26 \\
\hline
\multicolumn{9}{c}{ } \\
\multicolumn{9}{c}{\sedona} \\
\hline
Days to $B$-band max. & 15.7 & 16.6 & 16.6 & 16.0 & 15.2 & 17.2 & 17.3 & 16.6 \\
Si \textsc{ii} $\lambda6355$ velocity at max.\ [$\unit[10^3]{km \, s^{-1}}$]\tablenotemark{a} & 10.0 & 10.9 & 13.9 & 16.5 & 7.7 & 9.8 & 12.7 & 15.6 \\
Bolometric mag.\ at $\unit[+0]{d}$\tablenotemark{b} & $-17.8$  & $-18.5$ & $-19.1$ & $-19.4$ & $-17.2$ & $-18.2$ & $-19.0$ & $-19.4$ \\
Bolometric mag.\ at $\unit[+15]{d}$ & $-16.7$ & $-17.5$ & $-18.1$ & $-18.5$ & $-16.2$ & $-17.3$ & $-18.1$ & $-18.5$ \\
Bolometric mag.\ at $\unit[+30]{d}$ & $-15.8$ & $-16.5$  & $-17.2$ & $-17.6$ & $-15.4$ & $-16.4$ & $-17.1$ & $-17.6$ \\
$U$-band mag.\ at $\unit[+0]{d}$ & $-17.6$ & $-18.8$  & $-19.6$ & $-20.0$ & $-16.8$ & $-18.4$ & $-19.6$ & $-20.0$ \\
$U$-band mag.\ at $\unit[+15]{d}$ & $-15.2$ & $-16.8$ & $-17.5$ & $-18.9$ & $-13.7$ & $-16.3$ & $-18.0$ & $-18.8$ \\
$U$-band mag.\ at $\unit[+30]{d}$ & $-13.9$ & $-14.7$ & $-15.5$ & $-15.6$ & $-13.4$ & $-14.6$ & $-15.7$ & $-16.0$ \\
$B$-band mag.\ at $\unit[+0]{d}$ & $-17.8$ & $-18.7$ & $-19.3$ & $-19.6$ & $-17.1$ & $-18.4$ & $-19.2$ & $-19.6$ \\
$B$-band mag.\ at $\unit[+15]{d}$ & $-16.0$ & $-17.0$ & $-17.9$ & $-19.0$ & $-15.1$ & $-16.7$ & $-18.2$ & $-18.9$ \\
$B$-band mag.\ at $\unit[+30]{d}$ & $-15.2$ & $-15.9$ & $-15.9$ & $-16.3$ & $-14.8$ & $-15.7$ & $-15.9$ & $-16.5$ \\
$V$-band mag.\ at $\unit[+0]{d}$ & $-18.0$ & $-18.7$ & $-19.4$ & $-19.7$ & $-17.4$ & $-18.5$ & $-19.3$ & $-19.7$ \\
$V$-band mag.\ at $\unit[+15]{d}$ & $-17.0$ & $-17.9$ & $-18.7$ & $-18.9$ & $-16.3$ & $-17.6$ & $-18.6$ & $-18.9$ \\
$V$-band mag.\ at $\unit[+30]{d}$ & $-16.0$ & $-16.7$ & $-17.2$ & $-17.7$ & $-15.5$ & $-16.6$ & $-17.1$ & $-17.9$ \\
$R$-band mag.\ at $\unit[+0]{d}$ & $-18.2$ & $-18.8$ & $-19.3$ & $-19.5$ & $-17.6$ & $-18.5$ & $-19.2$ & $-19.4$ \\
$R$-band mag.\ at $\unit[+15]{d}$ & $-17.5$ & $-18.3$ & $-18.7$ & $-18.7$ & $-16.9$ & $-18.0$ & $-18.7$ & $-18.7$ \\
$R$-band mag.\ at $\unit[+30]{d}$ & $-16.4$ & $-17.0$ & $-17.6$ & $-18.2$ & $-15.9$ & $-17.0$ & $-17.6$ & $-18.3$ \\
$I$-band mag.\ at $\unit[+0]{d}$ & $-17.9$ & $-18.2$ & $-18.3$ & $-18.5$ & $-17.7$ & $-18.2$ & $-18.4$ & $-18.5$ \\
$I$-band mag.\ at $\unit[+15]{d}$ & $-17.3$ & $-17.5$ & $-18.3$ & $-17.9$ & $-17.1$ & $-17.9$ & $-17.7$ & $-17.9$ \\
$I$-band mag.\ at $\unit[+30]{d}$ & $-16.8$ & $-17.3$ & $-17.8$ & $-18.3$ & $-16.4$ & $-17.3$ & $-17.7$ & $-18.3$ \\
\hline
\multicolumn{9}{c}{ } \\
\multicolumn{9}{c}{\cmfgen} \\
\hline
Days to $B$-band max. & 14.8 & 16.3 & 16.3 & 15.3 & \nodata & \nodata & 17.2 & \nodata \\
Si \textsc{ii} $\lambda6355$ velocity at max.\ [$\unit[10^3]{km \, s^{-1}}$]\tablenotemark{a} & 10.7 & 11.9 & 13.6 & 14.9 & \nodata & \nodata & 12.1 & \nodata \\
Bolometric mag.\ at $\unit[+0]{d}$\tablenotemark{b} & $-17.6$  & $-18.4$  & $-19.0$ & $-19.4$ & \nodata& \nodata& $-19.0$ &\nodata \\
Bolometric mag.\ at $\unit[+15]{d}$ & $-16.8$  & $-17.4$  & $-18.0$ & $-18.5$ & \nodata& \nodata& $-18.0$ &\nodata \\
Bolometric mag.\ at $\unit[+30]{d}$ & \nodata  & $-16.6$  & $-17.3$ & $-17.7$ & \nodata& \nodata& $-17.3$ &\nodata \\
$U$-band mag.\ at $\unit[+0]{d}$ & $-17.1$  & $-18.7$  & $-19.6$ & $-20.0$ & \nodata & \nodata & $-19.5$ & \nodata \\
$U$-band mag.\ at $\unit[+15]{d}$ & $-15.6$  & $-16.7$  & $-18.1$ & $-19.0$ & \nodata & \nodata & $-17.9$ & \nodata \\
$U$-band mag.\ at $\unit[+30]{d}$ &\nodata  & $-15.9$  & $-17.1$ & $-18.0$ & \nodata & \nodata & $-16.9$ & \nodata \\
$B$-band mag.\ at $\unit[+0]{d}$ & $-17.6$ & $-18.5$ & $-19.2$ & $-19.6$ & \nodata & \nodata & $-19.1$ & \nodata \\
$B$-band mag.\ at $\unit[+15]{d}$ & $-15.8$ & $-16.9$ & $-18.1$ & $-18.7$ & \nodata & \nodata & $-17.9$ & \nodata \\
$B$-band mag.\ at $\unit[+30]{d}$ & \nodata & $-15.9$  & $-17.1$ & $-17.8$ & \nodata & \nodata & $-16.9$ & \nodata \\
$V$-band mag.\ at $\unit[+0]{d}$ & $-18.0$  & $-18.6$  & $-19.2$ & $-19.6$ & \nodata & \nodata & $-19.2$ & \nodata \\
$V$-band mag.\ at $\unit[+15]{d}$ & $-17.0$  & $-17.8$  & $-18.6$ & $-18.9$ & \nodata & \nodata & $-18.5$ & \nodata \\
$V$-band mag.\ at $\unit[+30]{d}$ &\nodata  & $-16.8$ & $-17.8$ & $-18.1$ & \nodata & \nodata & $-17.7$ & \nodata \\
$R$-band mag.\ at $\unit[+0]{d}$ & $-18.1$  & $-18.6$  & $-19.1$ & $-19.4$ & \nodata & \nodata & $-19.1$ & \nodata \\
$R$-band mag.\ at $\unit[+15]{d}$ & $-17.4$  & $-18.0$  & $-18.4$ & $-18.7$ & \nodata & \nodata & $-18.4$ & \nodata \\
$R$-band mag.\ at $\unit[+30]{d}$ &\nodata  & $-17.1$  & $-17.8$ & $-18.0$ & \nodata & \nodata & $-17.8$ & \nodata \\
$I$-band mag.\ at $\unit[+0]{d}$ & $-18.0$  & $-18.4$  & $-18.7$ & $-18.9$ & \nodata & \nodata & $-18.7$ & \nodata \\
$I$-band mag.\ at $\unit[+15]{d}$ & $-17.8$  & $-18.2$  & $-17.9$ & $-17.9$ & \nodata & \nodata & $-18.1$ & \nodata \\
$I$-band mag.\ at $\unit[+30]{d}$ &\nodata  & $-17.4$  & $-17.8$ & $-17.7$ & \nodata & \nodata & $-18.0$ & \nodata \\
\enddata
\tablenotetext{a}{Velocities are inferred from the wavelength of the absorption minimum.}
\tablenotetext{b}{Times are with respect to the time of $B$-band maximum.}
\end{deluxetable}




\begin{thebibliography}{}
\expandafter\ifx\csname natexlab\endcsname\relax\def\natexlab#1{#1}\fi
\providecommand{\url}[1]{\href{#1}{#1}}
\providecommand{\dodoi}[1]{doi:~\href{http://doi.org/#1}{\nolinkurl{#1}}}
\providecommand{\doeprint}[1]{\href{http://ascl.net/#1}{\nolinkurl{http://ascl.net/#1}}}
\providecommand{\doarXiv}[1]{\href{https://arxiv.org/abs/#1}{\nolinkurl{https://arxiv.org/abs/#1}}}

\bibitem[{{Baron} {et~al.}(1996){Baron}, {Hauschildt}, {Nugent}, \&
  {Branch}}]{baro96a}
{Baron}, E., {Hauschildt}, P.~H., {Nugent}, P., \& {Branch}, D. 1996, \mnras,
  283, 297, \dodoi{10.1093/mnras/283.1.297}

\bibitem[{{Bildsten} {et~al.}(2007){Bildsten}, {Shen}, {Weinberg}, \&
  {Nelemans}}]{bild07}
{Bildsten}, L., {Shen}, K.~J., {Weinberg}, N.~N., \& {Nelemans}, G. 2007,
  \apjl, 662, L95, \dodoi{10.1086/519489}

\bibitem[{{Blondin} {et~al.}(2018){Blondin}, {Dessart}, \& {Hillier}}]{blon18a}
{Blondin}, S., {Dessart}, L., \& {Hillier}, D.~J. 2018, \mnras, 474, 3931,
  \dodoi{10.1093/mnras/stx3058}

\bibitem[{{Blondin} {et~al.}(2017){Blondin}, {Dessart}, {Hillier}, \&
  {Khokhlov}}]{blon17a}
{Blondin}, S., {Dessart}, L., {Hillier}, D.~J., \& {Khokhlov}, A.~M. 2017,
  \mnras, 470, 157, \dodoi{10.1093/mnras/stw2492}

\bibitem[{{Colgate} \& {McKee}(1969)}]{cm69}
{Colgate}, S.~A., \& {McKee}, C. 1969, \apj, 157, 623, \dodoi{10.1086/150102}

\bibitem[{{Cyburt} {et~al.}(2010){Cyburt}, {Amthor}, {Ferguson}, {Meisel},
  {Smith}, {Warren}, {Heger}, {Hoffman}, {Rauscher}, {Sakharuk}, {Schatz},
  {Thielemann}, \& {Wiescher}}]{cybu10a}
{Cyburt}, R.~H., {Amthor}, A.~M., {Ferguson}, R., {et~al.} 2010, \apjs, 189,
  240, \dodoi{10.1088/0067-0049/189/1/240}

\bibitem[{{Dessart} {et~al.}(2014){Dessart}, {Hillier}, {Blondin}, \&
  {Khokhlov}}]{dess14b}
{Dessart}, L., {Hillier}, D.~J., {Blondin}, S., \& {Khokhlov}, A. 2014, \mnras,
  441, 3249, \dodoi{10.1093/mnras/stu789}

\bibitem[{{Fink} {et~al.}(2007){Fink}, {Hillebrandt}, \& {R{\"o}pke}}]{fhr07}
{Fink}, M., {Hillebrandt}, W., \& {R{\"o}pke}, F.~K. 2007, \aap, 476, 1133,
  \dodoi{10.1051/0004-6361:20078438}

\bibitem[{{Fryxell} {et~al.}(2000){Fryxell}, {Olson}, {Ricker}, {Timmes},
  {Zingale}, {Lamb}, {MacNeice}, {Rosner}, {Truran}, \& {Tufo}}]{fryx00}
{Fryxell}, B., {Olson}, K., {Ricker}, P., {et~al.} 2000, \apjs, 131, 273,
  \dodoi{10.1086/317361}

\bibitem[{{Ganeshalingam} {et~al.}(2010){Ganeshalingam}, {Li}, {Filippenko},
  {Anderson}, {Foster}, {Gates}, {Griffith}, {Grigsby}, {Joubert}, {Leja},
  {Lowe}, {Macomber}, {Pritchard}, {Thrasher}, \& {Winslow}}]{gane10a}
{Ganeshalingam}, M., {Li}, W., {Filippenko}, A.~V., {et~al.} 2010, \apjs, 190,
  418, \dodoi{10.1088/0067-0049/190/2/418}

\bibitem[{{Garnavich} {et~al.}(2004){Garnavich}, {Bonanos}, {Krisciunas},
  {Jha}, {Kirshner}, {Schlegel}, {Challis}, {Macri}, {Hatano}, {Branch},
  {Bothun}, \& {Freedman}}]{garn04}
{Garnavich}, P.~M., {Bonanos}, A.~Z., {Krisciunas}, K., {et~al.} 2004, \apj,
  613, 1120, \dodoi{10.1086/422986}

\bibitem[{{Giammichele} {et~al.}(2018){Giammichele}, {Charpinet}, {Fontaine},
  {Brassard}, {Green}, {Van Grootel}, {Bergeron}, {Zong}, \&
  {Dupret}}]{giam18a}
{Giammichele}, N., {Charpinet}, S., {Fontaine}, G., {et~al.} 2018, \nat, 554,
  73, \dodoi{10.1038/nature25136}

\bibitem[{{Goldstein} \& {Kasen}(2018)}]{gold18a}
{Goldstein}, D.~A., \& {Kasen}, D. 2018, \apjl, 852, L33,
  \dodoi{10.3847/2041-8213/aaa409}

\bibitem[{{Gronow} {et~al.}(2020){Gronow}, {Collins}, {Ohlmann}, {Pakmor},
  {Kromer}, {Seitenzahl}, {Sim}, \& {R{\"o}pke}}]{gron20a}
{Gronow}, S., {Collins}, C., {Ohlmann}, S.~T., {et~al.} 2020, \aap, 635, A169,
  \dodoi{10.1051/0004-6361/201936494}

\bibitem[{{Guillochon} {et~al.}(2010){Guillochon}, {Dan}, {Ramirez-Ruiz}, \&
  {Rosswog}}]{guil10}
{Guillochon}, J., {Dan}, M., {Ramirez-Ruiz}, E., \& {Rosswog}, S. 2010, \apjl,
  709, L64, \dodoi{10.1088/2041-8205/709/1/L64}

\bibitem[{{Guillochon} {et~al.}(2017){Guillochon}, {Parrent}, {Kelley}, \&
  {Margutti}}]{guil17a}
{Guillochon}, J., {Parrent}, J., {Kelley}, L.~Z., \& {Margutti}, R. 2017, \apj,
  835, 64, \dodoi{10.3847/1538-4357/835/1/64}

\bibitem[{{Hicken} {et~al.}(2009){Hicken}, {Challis}, {Jha}, {Kirshner},
  {Matheson}, {Modjaz}, {Rest}, {Wood-Vasey}, {Bakos}, {Barton}, {Berlind},
  {Bragg}, {Brice{\~n}o}, {Brown}, {Caldwell}, {Calkins}, {Cho}, {Ciupik},
  {Contreras}, {Dendy}, {Dosaj}, {Durham}, {Eriksen}, {Esquerdo}, {Everett},
  {Falco}, {Fernandez}, {Gaba}, {Garnavich}, {Graves}, {Green}, {Groner},
  {Hergenrother}, {Holman}, {Hradecky}, {Huchra}, {Hutchison}, {Jerius},
  {Jordan}, {Kilgard}, {Krauss}, {Luhman}, {Macri}, {Marrone}, {McDowell},
  {McIntosh}, {McNamara}, {Megeath}, {Mochejska}, {Munoz}, {Muzerolle},
  {Naranjo}, {Narayan}, {Pahre}, {Peters}, {Peterson}, {Rines}, {Ripman},
  {Roussanova}, {Schild}, {Sicilia-Aguilar}, {Sokoloski}, {Smalley}, {Smith},
  {Spahr}, {Stanek}, {Barmby}, {Blondin}, {Stubbs}, {Szentgyorgyi}, {Torres},
  {Vaz}, {Vikhlinin}, {Wang}, {Westover}, {Woods}, \& {Zhao}}]{hick09a}
{Hicken}, M., {Challis}, P., {Jha}, S., {et~al.} 2009, \apj, 700, 331,
  \dodoi{10.1088/0004-637X/700/1/331}

\bibitem[{{Hillier} \& {Dessart}(2012)}]{hill12a}
{Hillier}, D.~J., \& {Dessart}, L. 2012, \mnras, 424, 252,
  \dodoi{10.1111/j.1365-2966.2012.21192.x}

\bibitem[{{H{\"o}flich} {et~al.}(2017){H{\"o}flich}, {Hsiao}, {Ashall},
  {Burns}, {Diamond}, {Phillips}, {Sand}, {Stritzinger}, {Suntzeff},
  {Contreras}, {Krisciunas}, {Morrell}, \& {Wang}}]{hoef17a}
{H{\"o}flich}, P., {Hsiao}, E.~Y., {Ashall}, C., {et~al.} 2017, \apj, 846, 58,
  \dodoi{10.3847/1538-4357/aa84b2}

\bibitem[{Hunter(2007)}]{hunt07a}
Hunter, J.~D. 2007, Computing in Science \& Engineering, 9, 90,
  \dodoi{10.1109/MCSE.2007.55}

\bibitem[{{Jha} {et~al.}(2006){Jha}, {Kirshner}, {Challis}, {Garnavich},
  {Matheson}, {Soderberg}, {Graves}, {Hicken}, {Alves}, {Arce}, {Balog},
  {Barmby}, {Barton}, {Berlind}, {Bragg}, {Brice{\~n}o}, {Brown}, {Buckley},
  {Caldwell}, {Calkins}, {Carter}, {Concannon}, {Donnelly}, {Eriksen},
  {Fabricant}, {Falco}, {Fiore}, {Garcia}, {G{\'o}mez}, {Grogin}, {Groner},
  {Groot}, {Haisch}, {Hartmann}, {Hergenrother}, {Holman}, {Huchra},
  {Jayawardhana}, {Jerius}, {Kannappan}, {Kim}, {Kleyna}, {Kochanek},
  {Koranyi}, {Krockenberger}, {Lada}, {Luhman}, {Luu}, {Macri}, {Mader},
  {Mahdavi}, {Marengo}, {Marsden}, {McLeod}, {McNamara}, {Megeath}, {Moraru},
  {Mossman}, {Muench}, {Mu{\~n}oz}, {Muzerolle}, {Naranjo}, {Nelson-Patel},
  {Pahre}, {Patten}, {Peters}, {Peters}, {Raymond}, {Rines}, {Schild},
  {Sobczak}, {Spahr}, {Stauffer}, {Stefanik}, {Szentgyorgyi}, {Tollestrup},
  {V{\"a}is{\"a}nen}, {Vikhlinin}, {Wang}, {Willner}, {Wolk}, {Zajac}, {Zhao},
  \& {Stanek}}]{jha06b}
{Jha}, S., {Kirshner}, R.~P., {Challis}, P., {et~al.} 2006, \aj, 131, 527,
  \dodoi{10.1086/497989}

\bibitem[{{Jha} {et~al.}(2019){Jha}, {Maguire}, \& {Sullivan}}]{jha19a}
{Jha}, S.~W., {Maguire}, K., \& {Sullivan}, M. 2019, Nature Astronomy, 3, 706,
  \dodoi{10.1038/s41550-019-0858-0}

\bibitem[{{Karp} {et~al.}(1977){Karp}, {Lasher}, {Chan}, \&
  {Salpeter}}]{karp77a}
{Karp}, A.~H., {Lasher}, G., {Chan}, K.~L., \& {Salpeter}, E.~E. 1977, \apj,
  214, 161, \dodoi{10.1086/155241}

\bibitem[{{Kasen}(2006)}]{kase06}
{Kasen}, D. 2006, \apj, 649, 939, \dodoi{10.1086/506588}

\bibitem[{{Kasen} {et~al.}(2009){Kasen}, {R{\"o}pke}, \& {Woosley}}]{kase09b}
{Kasen}, D., {R{\"o}pke}, F.~K., \& {Woosley}, S.~E. 2009, \nat, 460, 869,
  \dodoi{10.1038/nature08256}

\bibitem[{{Kasen} {et~al.}(2006){Kasen}, {Thomas}, \& {Nugent}}]{ktn06}
{Kasen}, D., {Thomas}, R.~C., \& {Nugent}, P. 2006, \apj, 651, 366,
  \dodoi{10.1086/506190}

\bibitem[{{Kasen} \& {Woosley}(2007)}]{kase07a}
{Kasen}, D., \& {Woosley}, S.~E. 2007, \apj, 656, 661, \dodoi{10.1086/510375}

\bibitem[{{Kerzendorf} \& {Sim}(2014)}]{kerz14a}
{Kerzendorf}, W.~E., \& {Sim}, S.~A. 2014, \mnras, 440, 387,
  \dodoi{10.1093/mnras/stu055}

\bibitem[{{Kushnir} {et~al.}(2020){Kushnir}, {Wygoda}, \& {Sharon}}]{kush20a}
{Kushnir}, D., {Wygoda}, N., \& {Sharon}, A. 2020, \mnras, 499, 4725,
  \dodoi{10.1093/mnras/staa3017}

\bibitem[{{Maoz} {et~al.}(2014){Maoz}, {Mannucci}, \& {Nelemans}}]{maoz14a}
{Maoz}, D., {Mannucci}, F., \& {Nelemans}, G. 2014, \araa, 52, 107,
  \dodoi{10.1146/annurev-astro-082812-141031}

\bibitem[{{Miles} {et~al.}(2019){Miles}, {Townsley}, {Shen}, {Timmes}, \&
  {Moore}}]{mile19a}
{Miles}, B.~J., {Townsley}, D.~M., {Shen}, K.~J., {Timmes}, F.~X., \& {Moore},
  K. 2019, \apj, 871, 154, \dodoi{10.3847/1538-4357/aaf8a5}

\bibitem[{{Munari} {et~al.}(2013){Munari}, {Henden}, {Belligoli}, {Castellani},
  {Cherini}, {Righetti}, \& {Vagnozzi}}]{muna13b}
{Munari}, U., {Henden}, A., {Belligoli}, R., {et~al.} 2013, \na, 20, 30,
  \dodoi{10.1016/j.newast.2012.09.003}

\bibitem[{{Nomoto}(1982)}]{nomo82b}
{Nomoto}, K. 1982, \apj, 257, 780, \dodoi{10.1086/160031}

\bibitem[{{Nugent} {et~al.}(1997){Nugent}, {Baron}, {Branch}, {Fisher}, \&
  {Hauschildt}}]{nuge97}
{Nugent}, P., {Baron}, E., {Branch}, D., {Fisher}, A., \& {Hauschildt}, P.~H.
  1997, \apj, 485, 812, \dodoi{10.1086/304459}

\bibitem[{{Pakmor} {et~al.}(2013){Pakmor}, {Kromer}, {Taubenberger}, \&
  {Springel}}]{pakm13a}
{Pakmor}, R., {Kromer}, M., {Taubenberger}, S., \& {Springel}, V. 2013, \apjl,
  770, L8, \dodoi{10.1088/2041-8205/770/1/L8}

\bibitem[{{Pankey}(1962)}]{pank62a}
{Pankey}, Jr., T. 1962, PhD thesis, Howard University

\bibitem[{{Paxton} {et~al.}(2011){Paxton}, {Bildsten}, {Dotter}, {Herwig},
  {Lesaffre}, \& {Timmes}}]{paxt11}
{Paxton}, B., {Bildsten}, L., {Dotter}, A., {et~al.} 2011, \apjs, 192, 3,
  \dodoi{10.1088/0067-0049/192/1/3}

\bibitem[{{Pereira} {et~al.}(2013){Pereira}, {Thomas}, {Aldering}, {Antilogus},
  {Baltay}, {Benitez-Herrera}, {Bongard}, {Buton}, {Canto}, {Cellier-Holzem},
  {Chen}, {Childress}, {Chotard}, {Copin}, {Fakhouri}, {Fink}, {Fouchez},
  {Gangler}, {Guy}, {Hillebrandt}, {Hsiao}, {Kerschhaggl}, {Kowalski},
  {Kromer}, {Nordin}, {Nugent}, {Paech}, {Pain}, {P{\'e}contal}, {Perlmutter},
  {Rabinowitz}, {Rigault}, {Runge}, {Saunders}, {Smadja}, {Tao},
  {Taubenberger}, {Tilquin}, \& {Wu}}]{pere13a}
{Pereira}, R., {Thomas}, R.~C., {Aldering}, G., {et~al.} 2013, \aap, 554, A27,
  \dodoi{10.1051/0004-6361/201221008}

\bibitem[{{Perlmutter} {et~al.}(1999){Perlmutter}, {Aldering}, {Goldhaber},
  {Knop}, {Nugent}, {Castro}, {Deustua}, {Fabbro}, {Goobar}, {Groom}, {Hook},
  {Kim}, {Kim}, {Lee}, {Nunes}, {Pain}, {Pennypacker}, {Quimby}, {Lidman},
  {Ellis}, {Irwin}, {McMahon}, {Ruiz-Lapuente}, {Walton}, {Schaefer}, {Boyle},
  {Filippenko}, {Matheson}, {Fruchter}, {Panagia}, {Newberg}, {Couch}, \& {The
  Supernova Cosmology Project}}]{perl99}
{Perlmutter}, S., {Aldering}, G., {Goldhaber}, G., {et~al.} 1999, \apj, 517,
  565, \dodoi{10.1086/307221}

\bibitem[{{Phillips}(1993)}]{phil93a}
{Phillips}, M.~M. 1993, \apjl, 413, L105, \dodoi{10.1086/186970}

\bibitem[{{Polin} {et~al.}(2019){Polin}, {Nugent}, \& {Kasen}}]{poli19a}
{Polin}, A., {Nugent}, P., \& {Kasen}, D. 2019, \apj, 873, 84,
  \dodoi{10.3847/1538-4357/aafb6a}

\bibitem[{{Riess} {et~al.}(1998){Riess}, {Filippenko}, {Challis},
  {Clocchiatti}, {Diercks}, {Garnavich}, {Gilliland}, {Hogan}, {Jha},
  {Kirshner}, {Leibundgut}, {Phillips}, {Reiss}, {Schmidt}, {Schommer},
  {Smith}, {Spyromilio}, {Stubbs}, {Suntzeff}, \& {Tonry}}]{ries98}
{Riess}, A.~G., {Filippenko}, A.~V., {Challis}, P., {et~al.} 1998, \aj, 116,
  1009, \dodoi{10.1086/300499}

\bibitem[{{Shen} {et~al.}(2018{\natexlab{a}}){Shen}, {Kasen}, {Miles}, \&
  {Townsley}}]{shen18a}
{Shen}, K.~J., {Kasen}, D., {Miles}, B.~J., \& {Townsley}, D.~M.
  2018{\natexlab{a}}, \apj, 854, 52, \dodoi{10.3847/1538-4357/aaa8de}

\bibitem[{{Shen} \& {Moore}(2014)}]{shen14b}
{Shen}, K.~J., \& {Moore}, K. 2014, \apj, 797, 46,
  \dodoi{10.1088/0004-637X/797/1/46}

\bibitem[{{Shen} {et~al.}(2017){Shen}, {Toonen}, \& {Graur}}]{shen17c}
{Shen}, K.~J., {Toonen}, S., \& {Graur}, O. 2017, \apjl, 851, L50,
  \dodoi{10.3847/2041-8213/aaa015}

\bibitem[{{Shen} {et~al.}(2018{\natexlab{b}}){Shen}, {Boubert}, {G{\"a}nsicke},
  {Jha}, {Andrews}, {Chomiuk}, {Foley}, {Fraser}, {Gromadzki}, {Guillochon},
  {Kotze}, {Maguire}, {Siebert}, {Smith}, {Strader}, {Badenes}, {Kerzendorf},
  {Koester}, {Kromer}, {Miles}, {Pakmor}, {Schwab}, {Toloza}, {Toonen},
  {Townsley}, \& {Williams}}]{shen18b}
{Shen}, K.~J., {Boubert}, D., {G{\"a}nsicke}, B.~T., {et~al.}
  2018{\natexlab{b}}, \apj, 865, 15, \dodoi{10.3847/1538-4357/aad55b}

\bibitem[{{Sim} {et~al.}(2010){Sim}, {R{\"o}pke}, {Hillebrandt}, {Kromer},
  {Pakmor}, {Fink}, {Ruiter}, \& {Seitenzahl}}]{sim10}
{Sim}, S.~A., {R{\"o}pke}, F.~K., {Hillebrandt}, W., {et~al.} 2010, \apjl, 714,
  L52, \dodoi{10.1088/2041-8205/714/1/L52}

\bibitem[{{Sim} {et~al.}(2013){Sim}, {Seitenzahl}, {Kromer},
  {Ciaraldi-Schoolmann}, {R{\"o}pke}, {Fink}, {Hillebrandt}, {Pakmor},
  {Ruiter}, \& {Taubenberger}}]{sim13a}
{Sim}, S.~A., {Seitenzahl}, I.~R., {Kromer}, M., {et~al.} 2013, \mnras, 436,
  333, \dodoi{10.1093/mnras/stt1574}

\bibitem[{{Stritzinger}(2005)}]{stri05a}
{Stritzinger}, M.~D. 2005, PhD thesis, Technischen Universit\"{a}t M\"{u}nchen

\bibitem[{{Timmes} {et~al.}(1995){Timmes}, {Woosley}, \& {Weaver}}]{tww95}
{Timmes}, F.~X., {Woosley}, S.~E., \& {Weaver}, T.~A. 1995, \apjs, 98, 617,
  \dodoi{10.1086/192172}

\bibitem[{{Townsley} {et~al.}(2019){Townsley}, {Miles}, {Shen}, \&
  {Kasen}}]{town19a}
{Townsley}, D.~M., {Miles}, B.~J., {Shen}, K.~J., \& {Kasen}, D. 2019, \apjl,
  878, L38, \dodoi{10.3847/2041-8213/ab27cd}

\bibitem[{{Tsvetkov} {et~al.}(2013){Tsvetkov}, {Shugarov}, {Volkov},
  {Goranskij}, {Pavlyuk}, {Katysheva}, {Barsukova}, \& {Valeev}}]{tsve13a}
{Tsvetkov}, D.~Y., {Shugarov}, S.~Y., {Volkov}, I.~M., {et~al.} 2013,
  Contributions of the Astronomical Observatory Skalnate Pleso, 43, 94.
\newblock \doarXiv{1311.3484}

\bibitem[{{Wilk} {et~al.}(2020){Wilk}, {Hillier}, \& {Dessart}}]{wilk20a}
{Wilk}, K.~D., {Hillier}, D.~J., \& {Dessart}, L. 2020, \mnras, 494, 2221,
  \dodoi{10.1093/mnras/staa640}

\bibitem[{{Woosley} {et~al.}(1986){Woosley}, {Taam}, \& {Weaver}}]{wtw86}
{Woosley}, S.~E., {Taam}, R.~E., \& {Weaver}, T.~A. 1986, \apj, 301, 601,
  \dodoi{10.1086/163926}

\end{thebibliography}
\end{document}